# A Recursive Threshold Visual Cryptography Scheme

Abhishek Parakh and Subhash Kak
Department of Computer Science, Oklahoma State University
Stillwater, OK 74078

**Abstract:** This paper presents a recursive hiding scheme for 2 out of 3 secret sharing. In recursive hiding of secrets, the user encodes additional information about smaller secrets in the shares of a larger secret without an expansion in the size of the latter, thereby increasing the efficiency of secret sharing. We present applications of our proposed protocol to images as well as text.

**Keywords:** recursive hiding of secrets, visual cryptography, secret sharing, information efficiency.

**Introduction**

Conventional secret sharing schemes [1], which have several networking applications, are information theoretically very inefficient. For example, a ($k$, $n$) secret sharing scheme expands a secret of $b$ bits into $n$ shares each of at least $b$ bits in size. Furthermore, since only $k$ of these shares are needed to recreate the secret, each bit of any share, in a threshold secret sharing schemes, conveys at most $\lceil 1/k \rceil$ bits of the secret. If $k = n$, as in the case of a non-threshold scheme, where all the shares must be brought together to recreate the secret, the information conveyed by each bit of any share is $\lceil 1/n \rceil$ bits of the secret.

An extension of secret sharing schemes is visual cryptography that aims at splitting images into two or more shares such that when a predetermined number of shares are aligned and stacked together, the secret image is revealed [1],[3], without the requirement of any computation. However, conventional approaches to visual cryptography also suffer from inefficiency in terms of number of bits of secret conveyed per bit of shares.

Recursive hiding of secrets was proposed in [2], with applications to both images and text, to increase the efficiency of visual cryptography and to make it possible to incorporate additional secret information that serves as a steganographic channel. The idea involved is recursive hiding of smaller secrets in shares of larger secrets with secret sizes doubling at every step, thereby increasing the information that every bit of share conveys to $(n-1)/n$ bit of secret i.e. nearly 100%. However, the scheme described in [2] is a non-threshold scheme where all the shares are needed to recreate the secret.

In this paper, we extend the idea of recursive hiding of secrets to 2 out of 3 threshold scheme and apply it to both images and text. Further, the idea can be generalized to a 2 out of $n$ threshold scheme. However we deal with only binary images and regard each pixel as one bit of information, denoting black or white pixel.

**The proposed scheme**

For text represented as a binary sequence, a 2 out of 3 secret sharing scheme can be developed on a comparison based algorithm as follows: we divide the secret bit into 3 pieces $p_1$, $p_2$, and $p_3$ such that $p_1 = p_2 = p_3$ if we wish to encode bit 0, and $p_1 \neq p_2 \neq p_3$ if we wish to encode bit 1.

It is clear that to satisfy the above conditions we would need at least 3 symbols, say 0, 1 and 2. Therefore to encode bit 0 we could create pieces $p_1 p_2 p_3$ as 000, 111, or 222. Whereas the candidates to encode bit 1 would be 012 and all possible permutations of it, i.e. 021, 102, 120, 210, and 201. In all, to encode secret bit 0 and secret bit 1, we have 3 and 6 possibilities, respectively, out of which any one can be chosen to satisfy our requirement based on the secret encoded.

**Example:** If $M$ is a 27 bit long message that we wish to encode into 3 shares and the threshold is 2, then the shares $S_1$, $S_2$, and $S_3$ may be created as follows:

M : 011011010110110011100101101
$S_1$ : 102012012010201201201020102
$S_2$ : 110020022120111210101221001
$S_3$ : 121001002200021222001122200

Viewed as a ternary alphabet, the efficiency of this system is 33%. If 0,1 and 2 are encoded using prefix coding as 0, 10, and 11 respectively, then we are effectively mapping each bit of secret into 5 bits of shares and the efficiency is only $27/(27 \times 5) = 1/5$, i.e. 20%.

The above efficiency can be improved by recursively hiding secrets in the shares of $M$. However, since each bit is mapped into 3 shares, in order to take advantage of the recursive technique, the secrets at each step must increase by a factor of 3. We can then hide the following secrets $M_1$, $M_2$, and $M_3$ in $S_1$, $S_2$, and $S_3$ as follows:

| Secret | Shares | |
|---|---|---|
| $M_1$ : 1 | 0 | $S_{M_{11}}$ |
| | 2 | $S_{M_{12}}$ |
| | 1 | $S_{M_{13}}$ |
| $M_2$ : 010 | **0**11 | $S_{M_{21}}$ |
| | 0**2**1 | $S_{M_{22}}$ |
| | 00**1** | $S_{M_{23}}$ |
| $M_3$ : 110101101 | **011**122102 | $S_{M_{31}}$ |
| | 121**021**200 | $S_{M_{32}}$ |
| | 201220**001** | $S_{M_{33}}$ |

| $M$ : 011011010110110011100101101 | **011122102**011101221001121202 | $S_1$ |
|---|---|---|
| | 02010112**21210212000**101022100 | $S_2$ |
| | 002110112201211212**201220001** | $S_3$ |

<div align="center">Table 1. Recursive hiding of smaller messages in the shares of larger messages</div>

Note that at each step we have used the shares of the previous smaller messages to create the shares of larger messages, these smaller shares are denoted in bold. Also, we have distributed the shares at each step so that no player has access to all the shares of the smaller messages and hence, every message remains secure until at least two players come together. This approach is different from that discussed in [2], where the shares of smaller messages were all accumulated into one of the larger shares instead of distributing them among all the possible players. As a result, any player having that share which encodes the smaller images could in principle recreate these smaller images without the help of the other player, which in some cases might not be acceptable. Therefore, our new approach seems to be more secure for certain applications.

Also seen in table 1 is that using recursive hiding of secrets, we have been able to encode 13 characters of $M_1$, $M_2$, and $M_3$ and 27 characters of $M$ into shares of $M$ alone. As a result the efficiency is $(13+27)/(3 \times 27) = 40/81 \approx 1/2$. If one considers binary representations of each character then each share now conveys $(13+27)/(5 \times 27) = 8/27 \approx 1/3$ bits. Compared to 1/5 bits of conventional approaches, this is an almost 40% increase in efficiency.

**Recursive hiding and threshold visual cryptography**

The idea described in previous section can be applied to images to develop a recursive 2 out of 3 visual cryptographic scheme. For this purpose we divide each pixel into 3 subpixels as shown in table 2.

| Pixel | Partition 1 | Partition 2 | Partition 3 |
|---|---|---|---|
| 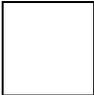 | 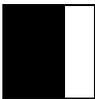 | 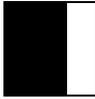 | 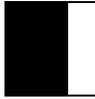 |
| | 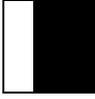 | 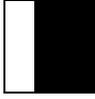 | 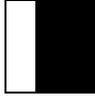 |
| | 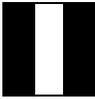 | 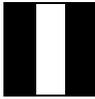 | 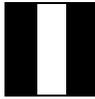 |

Table 2. Possible partitions for black and white pixels.

As seen in table 2, when the partitions of white pixel are stacked upon each other one third of the pixel is white and hence appears light gray to human eye. However, the subpixels of the black pixel are so arranged that when 2 shares are stacked together, the resulting pixel is completely dark.

Yet another way to create subpixels would be to have only one third of the subpixel colored dark. Therefore, when subpixels of a larger white pixel are stacked upon each other they would appear light gray and the stacking of the subpixels of a black pixel would result in dark gray. However, the human eye can perceives the difference between gray and completely dark pixels better than two different shades of gray itself. Hence our construction of subpixels in table 2.

We observe that the distribution of subpixels also corresponds to the comparison based splitting of secrets in the case of text. However, in this case, each subpixel can simply be represented by 1 bit. Therefore a white subpixel represents a 0 and black subpixel represents a 1. Their spatial distribution determines the manner in which they are stacked and the color they produce in the final image.

As an example to make the working of the proposed scheme clearer, we present in figure 1 the encoding of a 3x3 pixel image such that each share of the 3x3 image contains shares of a 1 pixel secret image and a 3x1 pixel secret image.

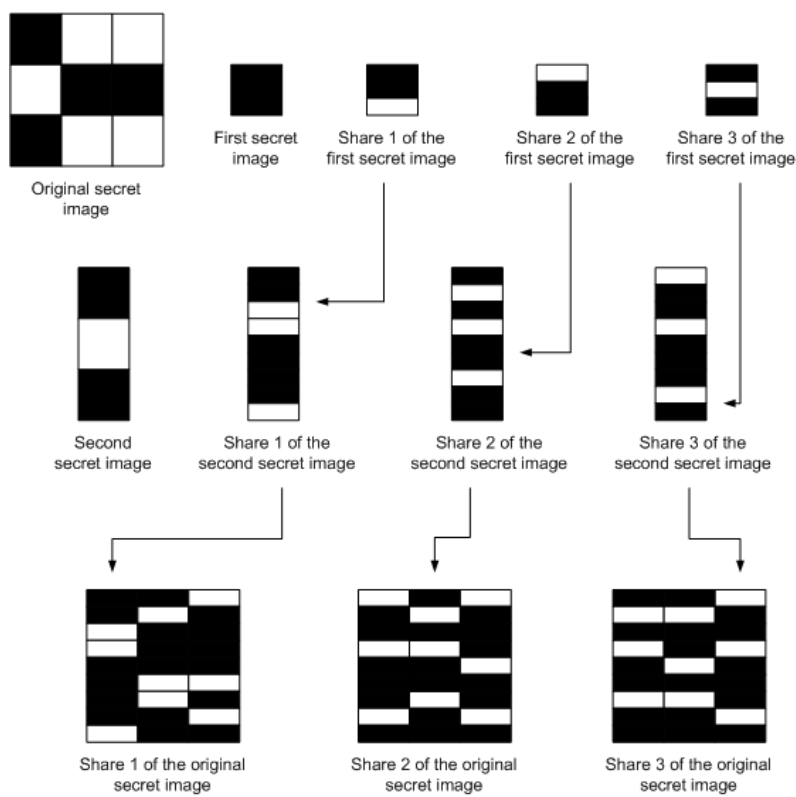

**Figure 1. Illustration of recursive hiding of secret images in the shares of larger original image using a 2 out of 3 threshold scheme.**

The subpixels of an original pixel can be represented as a matrix. For example if the original pixel was black then the 3 shares representing it can be written as $[100]^T$, $[010]^T$, and $[001]^T$. Since, these matrices can be stored as a sequence of bits; it implies that there is an expansion by a factor of 1×9=9 because the original black pixel can be represented as a single bit 1. If we were not to perform a recursive hiding, we would be creating 9×9=81 bits for each share corresponding to 9 pixels of the original image. However, using recursive hiding we have been able to hide additional 1×9+3×9=9+27=36 bits of information in those 81 bits, thereby increasing the information conveyed per share of the original image.

Higher efficiency could be achieved if we were to number the subpixels as 0, 1, and 2 and use prefix coding to represent these numbers and store them instead of storing the matrices or pixels. This would only lead to a per bit expansion factor of 5, instead of 9 and the efficiency improvement will be similar to that in the case of text, i.e. an improvement of 40%.

Figure 2 shows the application of the proposed scheme to three images, smallest image being a smiley face, next being a water mark and the third and the largest image being that of lena.

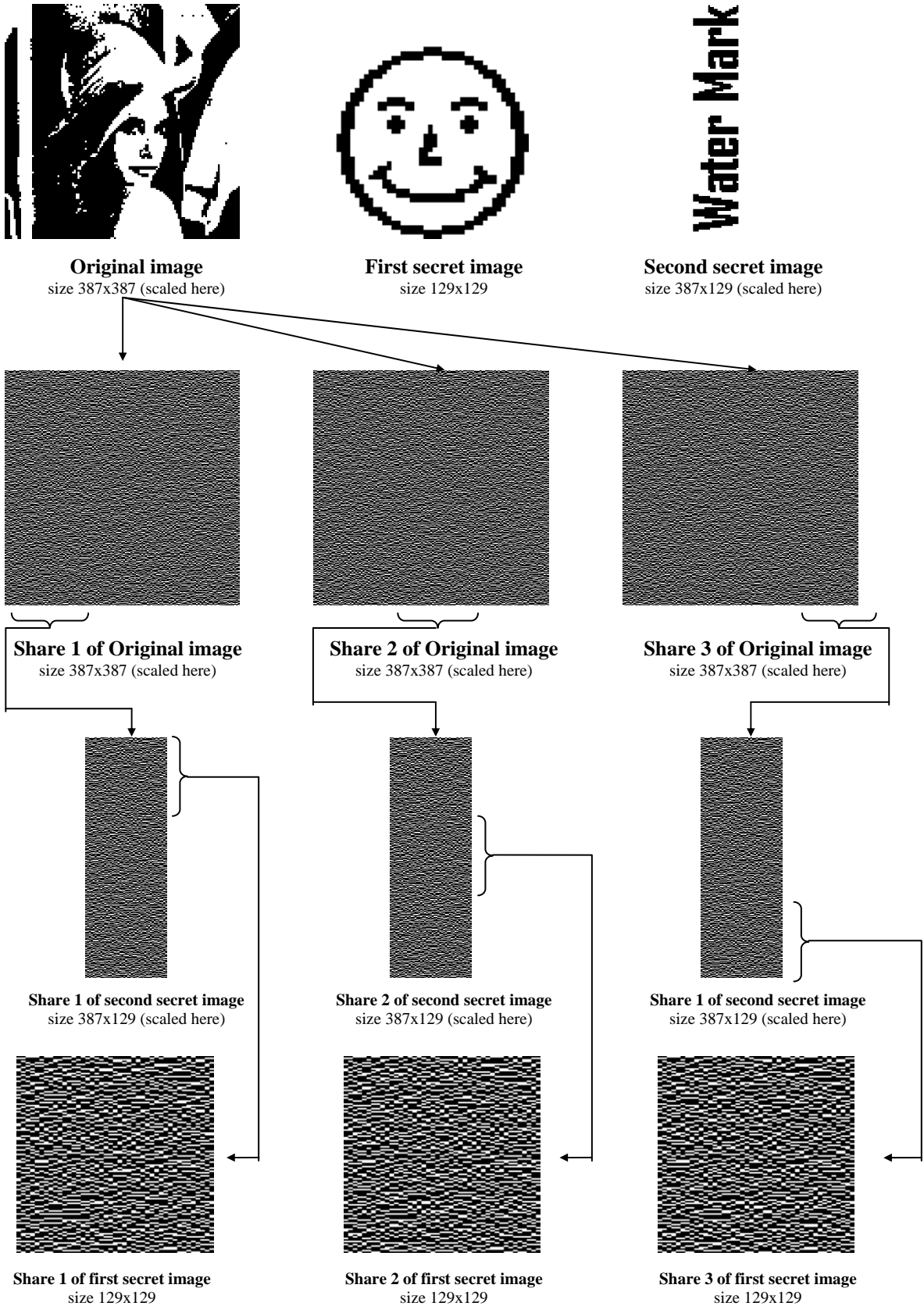

**Figure 2. Illustration of the process of recursive hiding of secrets in shares of larger original image**

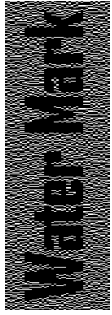 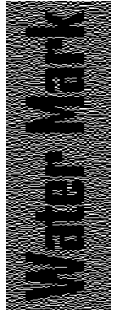 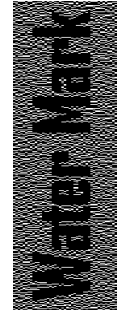

| **Regenerated second secret image from shares 1 and 2** size 387x129 (scaled here) | **Regenerated second secret image from shares 1 and 3** size 387x129 (scaled here) | **Regenerated second secret image from shares 2 and 3** size 387x129 (scaled here) |

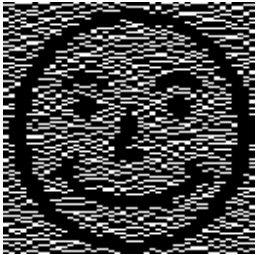 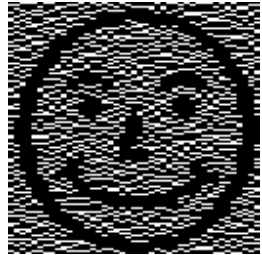 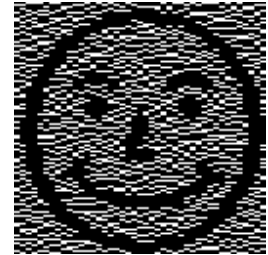

| **Regenerated first secret image from shares 1 and 2** size 129x129 | **Regenerated first secret image from shares 1 and 3** size 129x129 | **Regenerated first secret image from shares 2 and 3** size 129x129 |

**Figure 3. Illustration of regeneration of smaller images from the shares hidden inside the shares of the original larger image.**

## Conclusions

This paper presents a threshold scheme for recursive hiding of secrets. The construction presented is for a 2 out of 3 scheme; however it can be expanded to a 2 out of $n$ scheme as well. Recursive hiding serves as a steganographic channel that can be used to embed invisible watermarks, convey secret keys or encode authentication information. Further applications of such schemes would be in secure distributed storage of information over a network using secret sharing to enable storage of extra information in the shares, thereby decreasing network load and increasing efficiency.